\documentclass[10pt,onecolumn,letterpaper]{article}

\usepackage{iccv}
\usepackage{times}
\usepackage{epsfig}
\usepackage{graphicx}
\usepackage{amsmath}
\usepackage{amssymb}
\usepackage{algorithm}
\usepackage{algorithmic}
\usepackage{bbm}
\usepackage{amsfonts}
\usepackage{mathrsfs}
\usepackage{multirow}

\usepackage[pagebackref=true,breaklinks=true,colorlinks,bookmarks=false]{hyperref}

\iccvfinalcopy 

\def\iccvPaperID{406} 

\begin{document}

\title{Neural Network and Particle Filtering: A Hybrid Framework for Crack Propagation Prediction}

\author{Seyed Fouad Karimian\affilnum{1}, Ramin Moradi\affilnum{1}, Sergio Cofre-Martel\affilnum{1}, Katrina M. Groth\affilnum{1} and Mohammad Modarres\affilnum{1}}

\author{Seyed Fouad Karimian{$^{1}$},~ Ramin Moradi{$^{1}$},~ Sergio Cofre-Martel{$^{1}$},~ Katrina M. Groth{$^{1}$},\\~ Mohammad Modarres{$^{1}$}\\\vspace{-8pt}{\small~}\\
{$^{1}$}Center for Risk and Reliability, University of Maryland, MD, USA\\
{\small{Corresponding Author: \tt{foadkrmn@umd.edu}}}
}

\maketitle
\thispagestyle{empty}

\begin{abstract}
Crack detection, length estimation, and Remaining Useful Life (RUL) prediction are among the most studied topics in reliability engineering. Several research efforts have studied physics of failure (PoF) of different materials, along with data-driven approaches as an alternative to the traditional PoF studies. To bridge the gap between these two techniques, we propose a novel hybrid framework for fatigue crack length estimation and prediction. Physics-based modeling is performed on the fracture mechanics degradation data by estimating parameters of the Paris Law, including the associated uncertainties. Crack length estimations are inferred by feeding manually extracted features from ultrasonic signals to a Neural Network (NN). The crack length prediction is then performed using the Particle Filter (PF) approach, which takes the Paris Law as a move function and uses the NN’s output as observation to update the crack growth path. This hybrid framework combines machine learning, physics-based modeling, and Bayesian updating with promising results.\end{abstract}


\section{Introduction}

Although assumed to be identically manufactured, components always present some variability in their performance while in service. This variability can be seen in their degradation path and time to failure as they are tested under identical conditions. In engineering, fatigue is the most common degradation mechanism, and it has been under extensive study over the past century \cite{cui2002state}. The fluctuating loads applied to components result in damage accumulation. The damage starts at the micro-scale constituent levels of materials and grows to a macro-scale level, eventually causing a catastrophic failure. In metals, accumulated damage causes a macro-crack to form and develop until final fracture. The fatigue damage modeling and life prediction in metals encompass a rich history of research studies, mainly focused on damage mechanisms and damage accumulation models \cite{santecchia2016review}. Fatigue life prediction in metals is usually divided into two main regimes: crack initiation and crack propagation.
Fracture mechanics focuses on crack length estimation based on the applied stress and other factors such as geometry. One of the most widely accepted models in fracture mechanics is the Paris' Law \cite{paris1963critical}. The flexibility to adapt to unseen situations and the available historical information on physics-based models make them a reliable tool for fatigue life prediction. However, due to the stochastic nature of fatigue and the need to decrease the uncertainty in degradation prediction of sensitive structures, several data-driven methods have been developed over the past years. The implementation of data-driven methods seeks to exploit the interpretation of high volumes of data collected by sensors during the operational life of a component. Hence, the assessment of the health state can constantly be updated based on the operational conditions, allowing to guide the maintenance decision-making towards a condition-based one. 

One of the sensor measurements that have proven to be a good assessment tool for crack identification and propagation are ultrasonic signals, particularly Lamb waves. By studying anomalies in their behavior, such as dispersion characteristics and propagation speeds, Lamb waves can be used for damage identification \cite{he2013multi} and quantification through Bayesian updating \cite{yang2016probabilistic,chen2016research}. Other applications of Lamb wave properties for crack damage identification can be found in structural health monitoring through a defect imaging system \cite{muller2017structural}, an enhanced hierarchical probability damage-imaging \cite{li2018damage}, and on-line crack prognostics using particle filtering based method \cite{yuan2017line}. The compelling evidence of the strong relationship between ultrasonic waves and damage identification led researchers to apply Machine Learning (ML) approaches of high hierarchical order, aiming to extract abstract features from the signals \cite{lu2009artificial,aminpour2012applying,sbarufatti2014numerically}. Among these, Artificial Neural Networks (ANN) stand out as the most popular approach. Indeed, De Fenza et al. \cite{de2015application}  analyzed Lamb waves to determine the location and quantification of damage in metallic and composite plates, based on a damage index determined from the propagation of the waves. These indices were then used to train the ANN, which outputs the health state diagnostic of the plate. Sbarufatti et al. \cite{sbarufatti2014numerically} addressed a similar problem scenario, training an ANN with synthetically generated data, which was then validated with experimental data from a plate. In \cite{sbarufatti2014numerically}, indices were extracted from Lamb waves to create a feature vector, which was later used as input for two regression networks, one for localization and the second for quantification. Further, application of NN for damage identification and quantification based on wave signals have been applied in \cite{khan2015prognostics,nazarko2016damage,cofre2019deep}.\raggedbottom

\par Prediction accuracy in data-driven models heavily depends on the availability of reliable training data. Uncertainty in the remaining useful life (RUL) estimation of degradation processes increases when a limited amount of training data is available. Meanwhile, the adaptability of physics-based models to account for different conditions and availability of historical data can provide a reliable estimation of the degradation path specifically in the case of widely used materials. To incorporate the benefits of both data-driven and physics-based models, hybrid approaches are widely used to model the degradation behaviors \cite{neerukatti2014fatigue,loutas2017data,shahraki2017review}. Particle Filtering (PF) and its evolved versions (thoroughly discussed by Jouin et al. \cite{jouin2016particle}) are among the most popular hybrid prognostic techniques. PF can be updated in real-time, captures nonlinearity's, accounts for uncertainty, and can to perform prognostic and prediction tasks, which are great assets to hybrid techniques.

In this paper, we propose a novel hybrid Neural Network-Particle Filtering (NN-PF) model as a powerful probabilistic tool for crack length estimation and prognostics. This model fuses the physics-based degradation model for crack growth (Paris' Law) with the output of the data-driven approach (NN). To the best of our knowledge, there is only one similar application of an assembled neural network-particle filtering framework by Baraldi et al. \cite{baraldi2013ensemble}, in which the authors present a bagged ensemble artificial neural network to map the state of the particle to the measurement. Their main focus is to quantify the uncertainty over the RUL predictions. Our approach, however, focuses on the framework development, which assembles NN with PF to output a crack length estimation. 


\section{Proposed Framework}
A hybrid physics-based and data-driven framework is proposed to describe the stochastic nature of crack growth and decrease the uncertainty of crack propagation estimation. In this framework, fracture mechanics is used to model the physics of crack growth. The uncertainty in the component degradation path is captured through the physics-based model. Parameters of the governing equation of fatigue crack growth, Paris' Law, are estimated based on the available training data, through a probability distribution fitting. As a result, instead of describing the behavior of an unseen component using deterministic parameters, a series of potential crack growth paths with their associated probabilities are used.

\begin{figure*}
 \centering
 \includegraphics[height = 0.35\textwidth,width=1\textwidth]{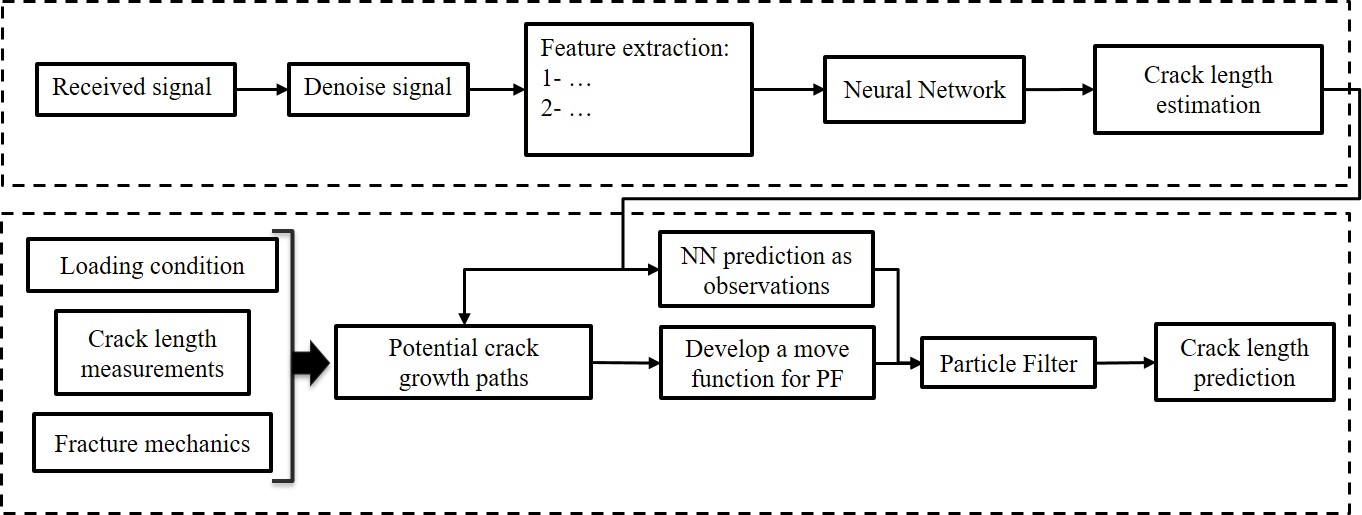}
 \caption{Flow diagram for the proposed framework.}
 \label{fig:framework}
\end{figure*}

Neural network (NN) estimates the crack length based on features extracted from ultrasonic signals recorded at different crack lengths. One important advantage of using NN for crack length estimation is that it is independent of the cycle numbers and loading conditions and only dependent on the signals. The crack length estimates from NN is used to select the most likely path of crack growth in the series of potential crack growth paths. Then, particle filter performs as a hybrid model, taking the selected crack growth path as the move function, and NN estimates as observations to update the degradation path of the fatigue test. This approach, while based on the general physics of crack growth, is tailored to the test conditions using NN outputs and is expected to decrease the uncertainty of RUL estimation and provide an accurate crack propagation estimation. Figure $\ref{fig:framework}$ shows the proposed framework. Details of each step are explained in the following. 

\subsection{Physics-Based Modeling}
The fatigue crack propagation prediction in this study is described by the linear elastic fracture mechanics (LEFM) introduced by Paris \cite{paris1963critical}. The Paris' Law equation explains the crack growth rate in a material as a function of its geometry and applied load, and it is defined as:
\begin{equation}
    \frac{da}{dN} = C(\Delta K)^m
    \label{Paris-law}
\end{equation}
In which $a$ is the crack length, $N$ is the number of cycles, and $da/dN$ is the crack growth rate with respect to the number of cycles. $C$ and $m$ are material dependent parameters, and $\Delta K$ is stress intensity factor defined as:
\begin{equation}
    \Delta K = f(g)\Delta\sigma\sqrt{\pi a}
    \label{delta-k}
\end{equation}
In Equation \ref{delta-k}, $f(g)$ is correction factor that depends on specimen and crack geometry, $\Delta \sigma$ is the applied stress and $a$ is the crack length. Equations \ref{Paris-law} and \ref{delta-k} can be used to describe the crack growth behavior of a known geometry and material component. Although the Paris' Law parameters for widely used materials are reported in many open-source databases, they are usually associated with high uncertainties. The material properties of a component can be measured by performing fatigue tests and crack length measurements and using:
\begin{equation}
    N_2 - N_1 = \int_{a_1}^{a_2}\frac{da}{C(f(g)\Delta\sigma\sqrt{\pi a})^m}
    \label{eqn:C m finding}
\end{equation}
To find the $C$ and $m$ parameters, crack growth data from the training database is fit to Equation \ref{eqn:C m finding}. The parameters are material dependent and are expected to yield identical values if multiple fatigue results are used from components of identical material. However, due to the uncertainties associated with tests, measurements, geometry, etc., values show a scattered behavior. After finding the Paris' Law parameters for each training test, a distribution over life at each crack length is fitted. A Weibull distribution is then used to capture the scatter in crack growth paths of all training test data. This, results in a distribution of crack growth paths describing the Paris' Law for the training set with its confidence intervals. In the case that the loading condition is changed from a constant amplitude to a variable amplitude, the change in Paris Law parameter $C$ is described using Newman's equation \cite{newman1984crack}.

\subsection{Deep Neural Network}
Neural networks have popularly been used to address problems regarding classification and regression for the last decades \cite{samarasinghe2016neural}. Given their hierarchical nature, NN could extract abstract information from a given input data $X_0$ by stacking multiple layers of nonlinear functions. A layer ($h_i$) is mathematically described as:
\begin{equation}
    h_{i} = \sigma(W_i^T \cdot X_{i-1} + b_{i}).
\end{equation}
Where, $W_i$ and $b_i$ are the so-called weights and bias of layer $i$, respectively. $X_{i-1}$ is the input data to the layer, and $\sigma$ is known as the activation function, which is a nonlinear function such as the hyperbolic tangent (tanh), rectifier linear unit (ReLU), sigmoid, etc. Each of these nonlinear transformations are known as layers. The number of outputs in each layer (i.e., $h_i$'s dimension) is equal to the number of neurons or hidden units, that the layer has. For a multi-layer neural network (i.e., Deep Neural Network), a layer's outputs $h_i$ correspond to the input data $X_{i+1}$ of the next layer. The output of the whole network ($y$) is then given by:
\begin{equation}
    y = W^T_l \cdot h_l + b_l
\end{equation}
where $l$ corresponds to the number of hidden layers of the network.


The neural network used in this framework has two hidden layers of ten hidden units each, with ReLU as the activation function. The network takes as input different features extracted from the Lamb Waves signals, and it outputs the estimated crack length corresponding to those features. Note that the network does not take the number of cycles as an input. Thus, crack estimation is solely based on the abstract interpretation of the Lamb waves. 

The training cost for the network is set as a combination of the prediction error and a penalization function $T$ (Equation \ref{eq: T}). This function was introduced as a penalization for the PHM Data Challenge 2019 \cite{2019PHMC82:online}. This gives a greater penalization for prediction errors at greater crack lengths. The training cost function is then defined as: 

\begin{equation}
    \Delta = y_i-W^T\cdot X_i
\end{equation}
\begin{equation}
    T = 2.0 + 10.0 \cdot y_i 
    \label{eq: T}
\end{equation}
\begin{equation}
    Cost =  \frac{1}{N}\sum_{i=1}^N \Delta^T \cdot T
\end{equation}
The network is trained using stochastic gradient descent through back-propagation with Adam optimizer \cite{kingma2014adam}. The learning rate was set to be 0.001. Further, the training of the network does not depend on the particle filtering process, and we only train the network using a random sample set from the given dataset. Hence, the output corresponds to one output per input data, rather than a distribution as in \cite{baraldi2013ensemble}. The network was trained using Python 3.6 through Tensorflow 1.13 in Windows 10. An Intel i7 9700k CPU, a Titan RTX GPU of 24GB, and 32GB of RAM were used as the hardware.

\subsection{Particle Filtering}
Particle Filtering (PF) is a sequential Monte Carlo-based computational technique that is frequently used for Bayesian prognostics of nonlinear and/or non-Gaussian processes.
Considering a system whose state at the discrete time step $t_k=k\Delta{t}$ is represented by a vector $x_k$. The state space model can describe the system's evolution as: 
\begin{equation}
    x_k = f_k(x_{k-1}, \omega_{k-1})
\end{equation}
Where $f_k$ is the nonlinear state transition function (aka move function) and $\omega$ is the state noise vector. Noise is being introduced in the modeling to account for the stochasticity involved in fatigue crack growth that can vary from specimen to specimen. The objective in the particle filtering process is to recursively estimate $x_k$ from measurements $z_k$ which can be described by Equation \ref{eqn:noise}.

\begin{equation}
\label{eqn:noise}
    z_k = h_k(x_k, \nu_k)
\end{equation}

Where $h_k$ is the measurement function and $\nu_k$ is the measurement noise. The Bayesian approach uses the following probability density function (PDF) $p(x_k{\mid}z_{1:k})$ to estimate the dynamic state $x_k$, given the measurements $z_k$ up to time $k$. This PDF is calculated recursively from $t_1$ to $t_k$ using Equation \ref{eqn:Bayes}. 

\begin{equation}
\label{eqn:Bayes}
    p(x_k \vert z_{1:k})= \frac{p(z_k \vert x_k). p(x_k \vert z_{1:k-1})}{p(z_k \vert z_{1:k-1})}
\end{equation}

To numerically perform PF, one should assume that a set of random samples, i.e., particles, $x_{k-1}^i, i=1,. . .,N$ of the system state at time $k-1$ are available as a realization for the posterior $p(x_{k-1} \vert z_{1:k-1})$. Prediction step at time $k$ is accomplished by: 
\begin{itemize}
    \item Sampling from the probability distribution of the system noise $\omega_{k-1}$
    \item Simulating the system dynamics (application of the move function) to generate a new set of samples $x_k^i$ which are realizations of the predicted probability distribution $p(x_k \vert z_{1:k-1})$. 
    \item Updating each sampled particle's $x_k^i$ assigned weight $w_k^i$ based on the likelihoods of the observations $z_k$ collected at time $k$. An approximation of the posterior distribution $p(x_k \vert z_{1:k})$ can then be obtained from the weighted samples ($x_k^i$,$w_k^i$), $i=1,\dots,N$. 
\end{itemize}
    

In this framework the move function for PF is determined by the physics-based modeling. The initial particles distribution mean and standard deviation is determined according to the upper bound and lower bound (i.e., 5\% and 95\% confidence) of the crack growth path obtained by physics-based modeling at a reasonably low cycle number (in which we are confident that the component is healthy). The impact of the observations is determined by the standard deviation of the distribution, whose mean is placed on the observation point and the particles weights are updated by it. This standard deviation value, along with the noise of the move function, is optimized in a way to achieve maximum prediction accuracy on the training data. To do so, it is considered that NN predictions are only available for the first few cycles in training tests. Then, using PF, the crack lengths in higher cycle numbers are predicted.

\section{Case Study}

\subsection{Data description}
The dataset used for this study is publicly available at the PHM data challenge 2019 competition \cite{2019PHMC82:online}. The dataset describes fatigue tests on lap-shear joints made of aluminum alloy, reporting ultrasonic signals and crack length measurements. The database includes six training tests and two validation tests. In the training tests, for each specimen, a set of cycle numbers and their corresponding crack lengths are provided. For each crack length, two sets of ultrasonic signals are given, which include an actuation and a received signal. The goal is to determine the crack lengths at given cycle numbers for two validation tests. In the validation data, ultrasonic signals are only provided for the first few cycles, and prognostics needs to be carried out for a determined number of cycles with no provided signals. 

\begin{figure}[h]
    \centering
    \includegraphics[width=0.5\linewidth]{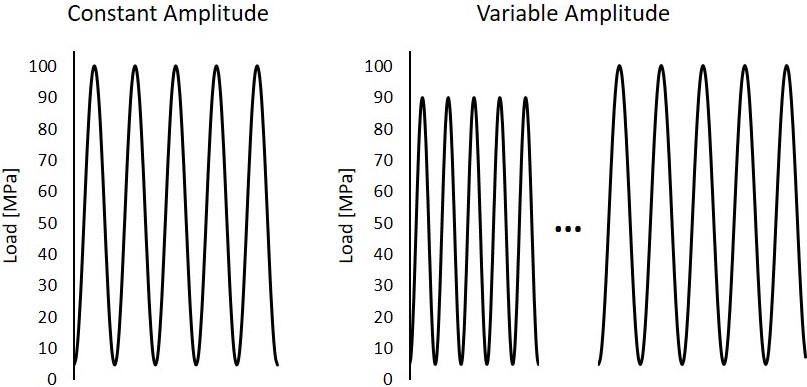}
    \caption{Tests loading conditions.}
    \label{fig:loading condition}
\end{figure}

All training tests and one validation test are fatigued with a constant amplitude loading, while the second validation test is fatigued under variable amplitude loading. Constant amplitude loading is described with a maximum stress $\sigma_{max} = 100.21\, \text{MPa}$ and a minimum stress $\sigma_{min} = 4.77\, \text{MPa}$, at a $5 \, \text{Hz}$ frequency. Variable amplitude spectrum is described as $500$ cycles with $\sigma_{max} = 90\, \text{MPa}$ and $\sigma_{min} = 4.77\, \text{MPa}$ followed by $500$ cycles with $\sigma_{max} = 100.21\, \text{MPa}$ and $\sigma_{min} = 4.77\, \text{MPa}$, at the same $5 \, \text{Hz}$ frequency. The loading conditions are shown in Figure \ref{fig:loading condition}. The only information available about the geometry of the specimen is the distance between actuator and receiver sensors that is $161.0\, \text{mm}$.

\subsection{Crack growth path}
The parameters of the Paris' Law are evaluated for each training test separately. The results of all training tests are used to find the mean and confidence intervals of the crack growth path for the specimen and material. Since the material type is not known, no prior distribution is used for the Paris' Law parameters. Furthermore, the geometry coefficient in Equation \ref{delta-k} cannot be directly selected due to the lack of knowledge about the geometry. Therefore, based on the description of the problem which states the crack is bracketed by the sensors, an equivalent center-cracked plate with tension loading is selected. The geometry coefficient, in this case, is defined as:

\begin{equation}
    f(g) = \sqrt{sec \Big(\frac{\pi a}{2b}\Big)}
    \label{f(g)}
\end{equation}

\begin{figure}[h]
\centering
 \includegraphics[width=0.5\linewidth]{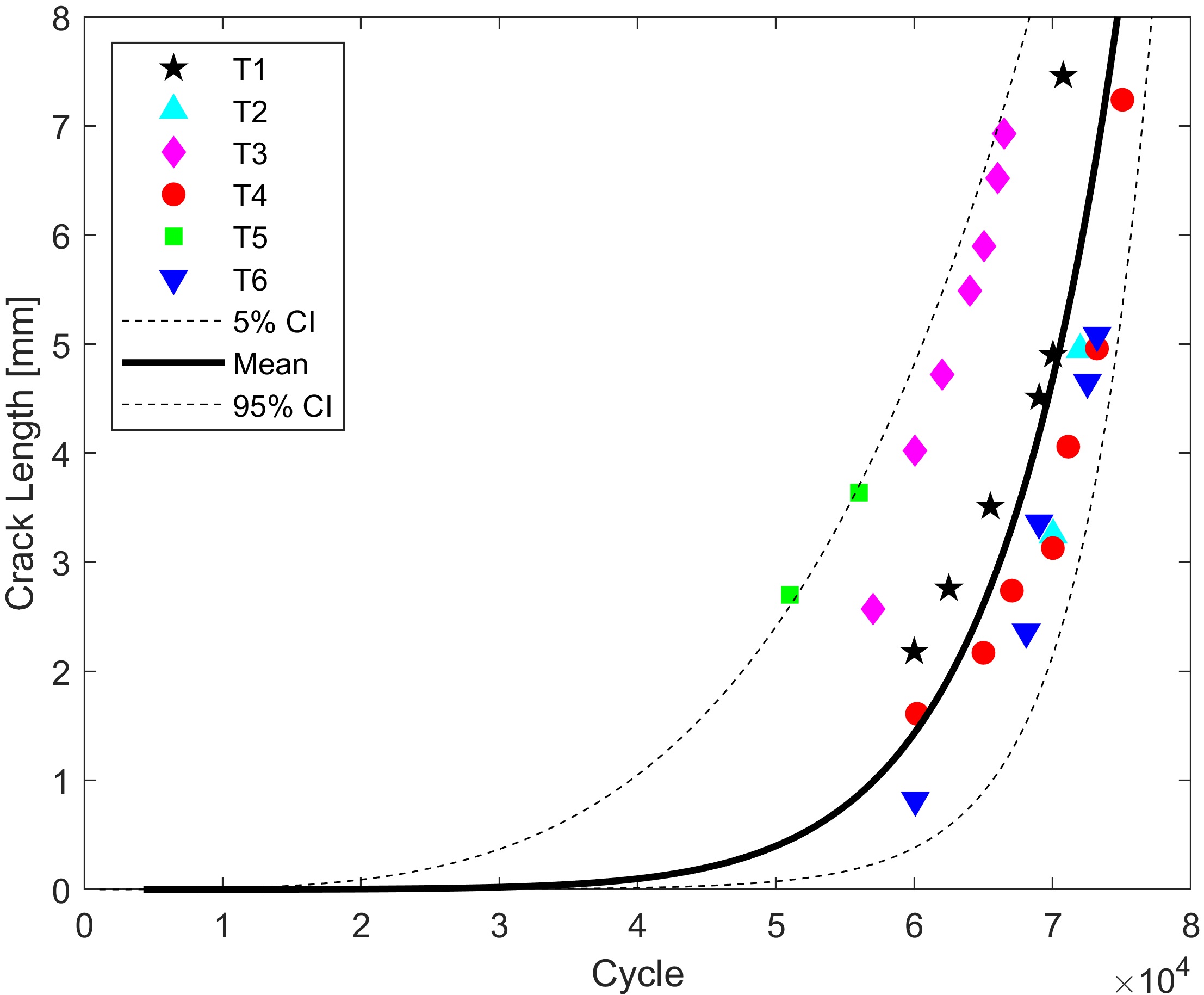}
 \caption{Crack growth path distribution based on the training data.}
 \label{fig:crack growth path}
\end{figure}

Where $2a$ is the crack length and $2b$ is the specimen width. The use of this $f(g)$ requires information about the specimen width. Hence, an optimization on the parameter $b$ of Equation \ref{f(g)} and parameters $C$ and $m$ of Equation $\ref{Paris-law}$ is performed using Equation \ref{eqn:C m finding}. The objective of the optimization is to minimize the error in crack length estimation using the known crack lengths. The optimization is done on all of the training tests individually. This results in a series of six values for the specimen width (parameter $b$). All the widths found are similar with an average of $39 mm$ and a standard deviation of $3 mm$. The optimization of Paris' Law parameters is then repeated for each training test using the average value for specimen width. Since the actual geometry of the specimen is different from the one assumed, the actual specimen width is also expected to be different from the equivalent specimen width found. The crack growth path of the training tests are used to find the mean and the confidence intervals for the crack growth path of the material and geometry. The mean, 5\% and 95\% confidence bounds of the crack growth path are found by fitting a Weibull distribution over the cycle numbers at different crack lengths. The results are shown in Figure \ref{fig:crack growth path}, where we can see that the mean and the confidence interval capture all the variability in the crack growth paths for the different training tests.

\subsection{Data cleaning}
The first step to use the ultrasonic signals is denoising. Since no information is provided regarding the actuation signals, we perform denoising aimed to increase the correlation coefficient for each set of two signals in the training tests. In order to increase the accuracy of signal denoising, only the signals that correspond to zero crack length are used. The correlation coefficient in each pair of signals is computed for both actuation and received signals.

\begin{figure}[h]
\centering
 \includegraphics[width=1\linewidth]{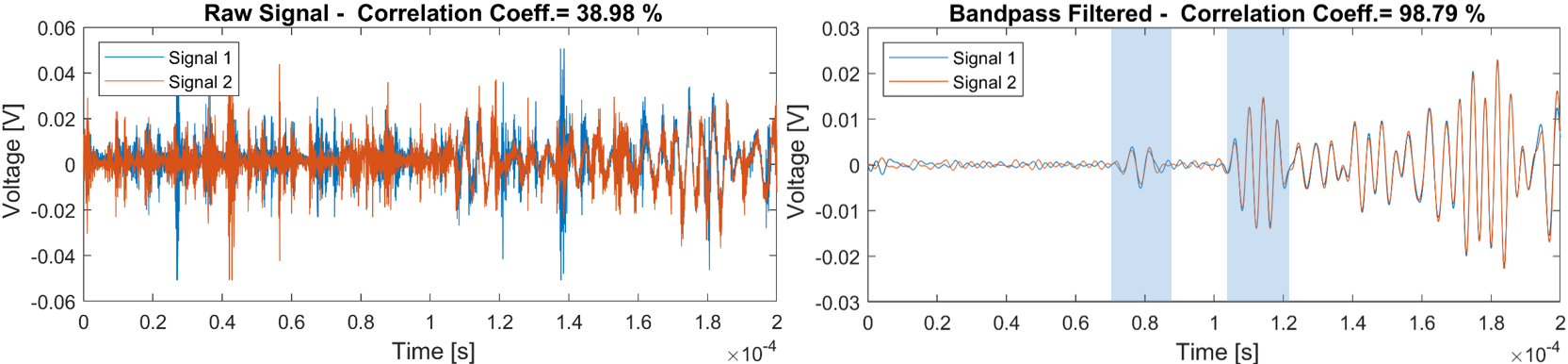}
 \caption{Raw signal vs. bandpass filtered signal and the time windows selected.}
 \label{fig:signalDenoise}
\end{figure}

Two methods of signal denoising are used in this study to achieve higher signal-to-noise ratios. The first one is wavelet analysis, where a 'db10' wavelet is used to denoise signals. The method is selected following the analysis presented in \cite{li2018damage}, which showed to be an effective denoising approach. The second method is based on the power spectrum of the signals. For each signal, the power spectrum shows a peak in the lower frequencies followed by an oscillation around lower power values for the rest of the frequency spectrum. Hence, the range of frequency showing the highest power spectrum is selected and used as a bandpass filter to denoise signals. It is assumed that the actual signal has the highest values in the power spectrum and therefore, the corresponding frequencies can be used to denoise signals. The average correlation coefficient for all the signals with zero crack length are summarized in Table $\ref{denoise summary}$. It can be seen that the bandpass filter has the highest value. Figure $\ref{fig:signalDenoise}$ shows a set of received signals before and after bandpass filtering. We observe that while the bandpass filtered signals have the same shape, raw signals include lots of noise, which significantly decreases the correlation coefficient. 

\begin{table}[hb]
\centering
\caption{Comparison of two different signal denoising methods.}
    \begin{tabular}{ |p{3cm}||p{2cm}|p{2cm}|  }
    \hline
     \multicolumn{3}{|c|}{Mean Correlation Coefficient} \\
    \hline
    Signal& Actuation & Received\\
    \hline
    Raw    & 96.31\% & 81.01\% \\
    Wavelet filter & 98.16\% & 83.48\% \\
    \textbf{bandpass filter}    & \textbf{99.83}\% & \textbf{97.29}\% \\
    \hline
    \end{tabular}
\label{denoise summary}
\end{table}

\subsection{Signal feature extraction}
In the simplest approach, one can analyze the whole recorded signal and calculate commonly used features like phase change \cite{yang2016probabilistic}, correlation coefficient \cite{li2018damage}, information entropy \cite{KARIMIAN2020106771}, normalized amplitude \cite{sun2017lamb}, normalized energy \cite{wang2018model}, and time of flight \cite{li2018damage}. Doing so would introduce a considerable amount of noise and uncertainty to the results, since the recorded signals are contaminated by boundary waves reflections and external sources of noise. In the literature, a specific portion of the signals is often considered for further analysis based on the expected time of receiving the actuation signal. This time window is determined by knowing the geometry and material of the test specimens and time of flight diffraction (ToFD) techniques \cite{sinclair2010enhancement,nath2010reliability}. However, for our case study the exact geometry, material and piezoelectric sensors specifications are not known to perform a thorough ToFD. 

Considering the above, to select the most important portion of the signal, a time window with the length of the actuation signal is considered. This window is then moved along the received signals to find the most correlated portions of the received signals compared to the actuation signal. The two most correlated portions approximate time windows are shown in Figure \ref{fig:signalDenoise}.  

\begin{figure}[h]
\centering
 \includegraphics[width=0.7\linewidth]{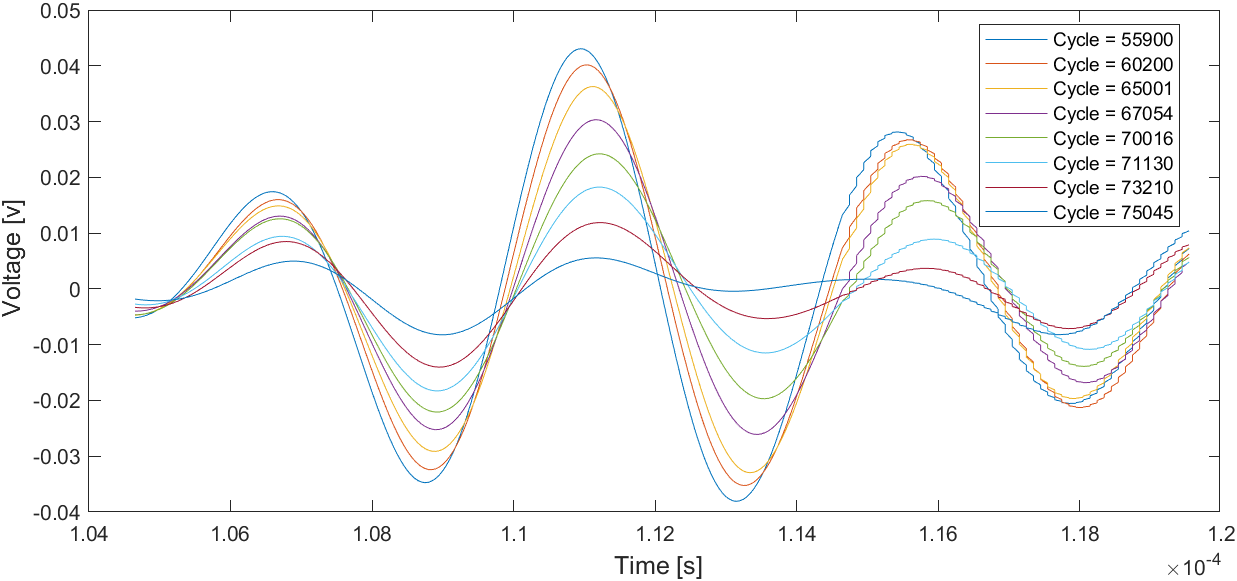}
 \caption{Signals at different cycles - Test 4.}
 \label{fig:Signal_window}
\end{figure}

This approach is justified since the actuation lamb wave frequency would remain constant when propagating through the material and only its energy (amplitude) gets dissipated. Further analysis of the extracted signals revealed that the features from signals corresponding to the second window (later in time) correlate better with the crack lengths. Therefore, only the second time window is used for further analysis. Figure $\ref{fig:Signal_window}$ shows the signals extracted from the second window in test 4, where we can see a change in the trend of the signals corresponding to different cycle numbers (i.e., different crack lengths). 

Features investigated in this study are: 
\begin{enumerate}
    \item \textbf{Pearson Correlation Coefficient} which is a measure of the linear dependence between two random variables. If there are $N$ observations for each signal, then the coefficient is calculated as: 
    \begin{equation} 
        \rho(A,B) = \frac{1}{N-1}\sum_{i=1}^N(\overline{\frac{A_i-\mu_B}{\sigma_A}})(\frac{B_i-\mu_B}{\sigma_B})
    \end{equation}
    Where $\mu$ and $\sigma$ are the mean and standard deviation of random variables $A$ and $B$, respectively. In this study, $A$ and $B$ are the received signals at different crack lengths for each test.

\item \textbf{Phase change} between different measurements of each test specimen \cite{yang2016probabilistic}. Phase change between two signals can be calculated with different methods. In this study, we use the Discrete Fourier Transform (DFT) and Maximum Likelihood estimation (MLE) of the properties of the signals. The phase difference between harmonic components of lamb wave signals is found as the phase difference of the harmonics DFT phase spectrum values of Lamb wave signals. A detailed mathematical explanation for various phase change calculation methods is provided by Sedlacek and Krumpholc \cite{sedlacek2005digital}. The phase change relative to the received signal when there is no crack is considered as a feature in this paper. 

\item \textbf{Energy} of the received signals. Cracks, damages, or any  anomalies in the structure of the material would cause energy dissipation when the signal is traveling through the material. The energy of the actuation signal is considered as the total amount of energy released into the material. Thus, the energy of the received signals are calculated as a percentage of the actuation signal's energy. Energy of a signal calculated as follows. 

\begin{equation}
    E_s = \int_{-\infty}^{\infty} |x(t)|^2 dt
    \label{eqn:signalEnergy}
\end{equation}

\item \textbf{Information Entropy}. This feature has recently received attention as a new parameter to quantify fatigue damage using recorded signals \cite{KARIMIAN2020106771}. Information entropy relies on the information content of the distribution of received signals. It uses the signal distribution characteristics to measure the information content of the signal and consequently quantifies the damage. The information entropy defined by Shannon \cite{shannon1948mathematical} is expressed as:
\begin{equation}
    I = -c\sum_{i=1}^{n}P(x_i)ln(1/P(x_i))
    \label{eqn:Shannon}
\end{equation}

 \begin{figure}[h]
     \centering
     \includegraphics[width=1\linewidth]{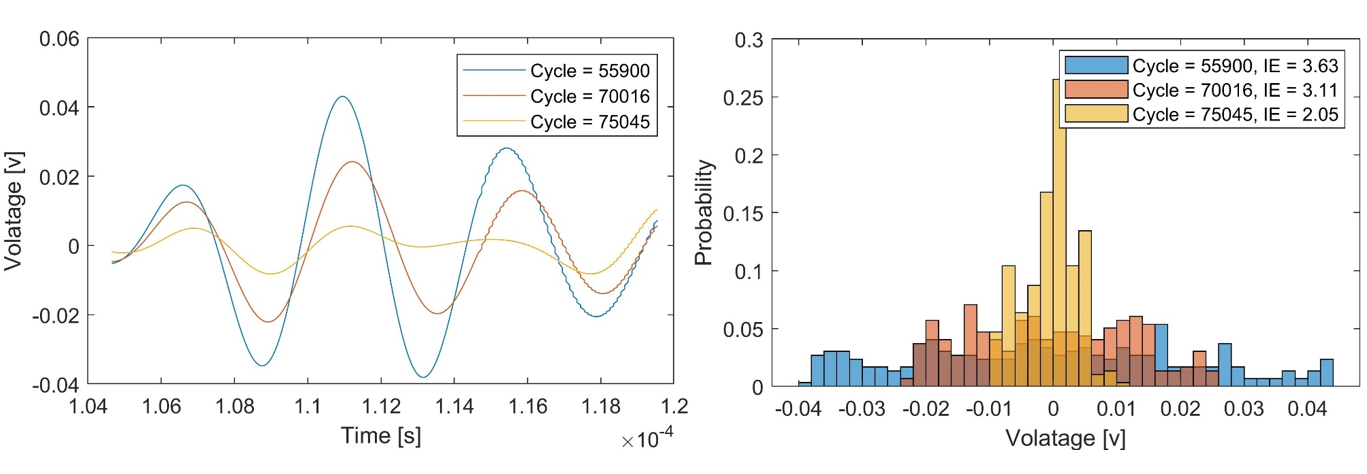}
     \caption{Signals, their voltage histograms and their respective information entropy (IE) values.}
     \label{fig:sign_hist}
 \end{figure}
 

Where $I$ is the information entropy, $c$ is a constant considered to be unity in this study and  $P(x_i)$ is the probability distribution of the random variable. In the case of ultrasonic signals, voltage is the random variable and the histogram of signals is used as a non-parametric probability distribution. Figure $\ref{fig:sign_hist}$ shows how the change in signal at different cycles results in a change of histogram and consequently its information entropy. Figure \ref{fig:train_entr} shows the trend of information entropy change in all tests. 

\begin{figure}[h]
    \centering
    \includegraphics[width=0.45\linewidth]{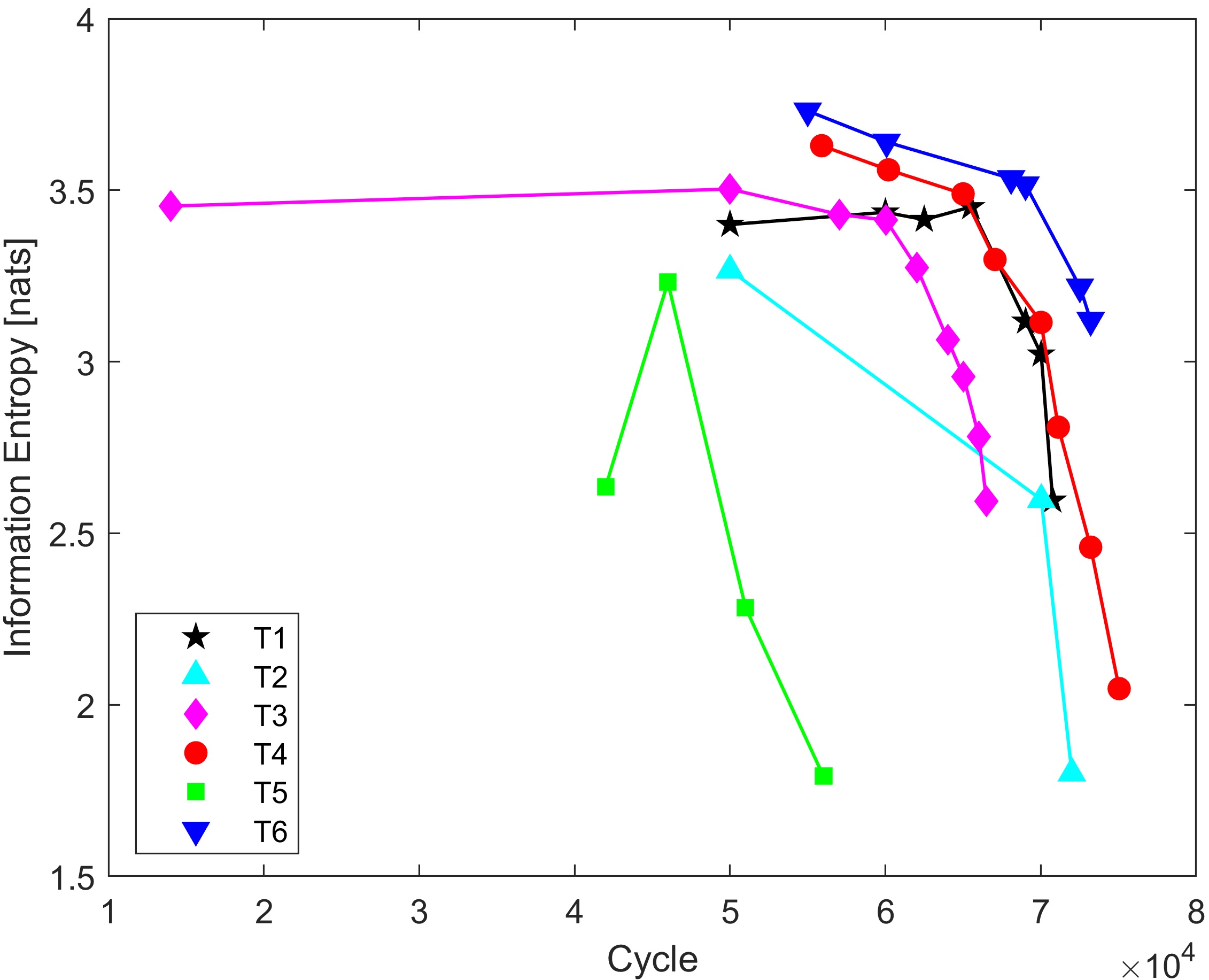}
    \caption{Trend of change in information entropy of training set.}
    \label{fig:train_entr}
\end{figure}


\end{enumerate}

\subsection{Results and discussion}

\begin{figure*}[hb]
    \centering
    \includegraphics[width=1\textwidth]{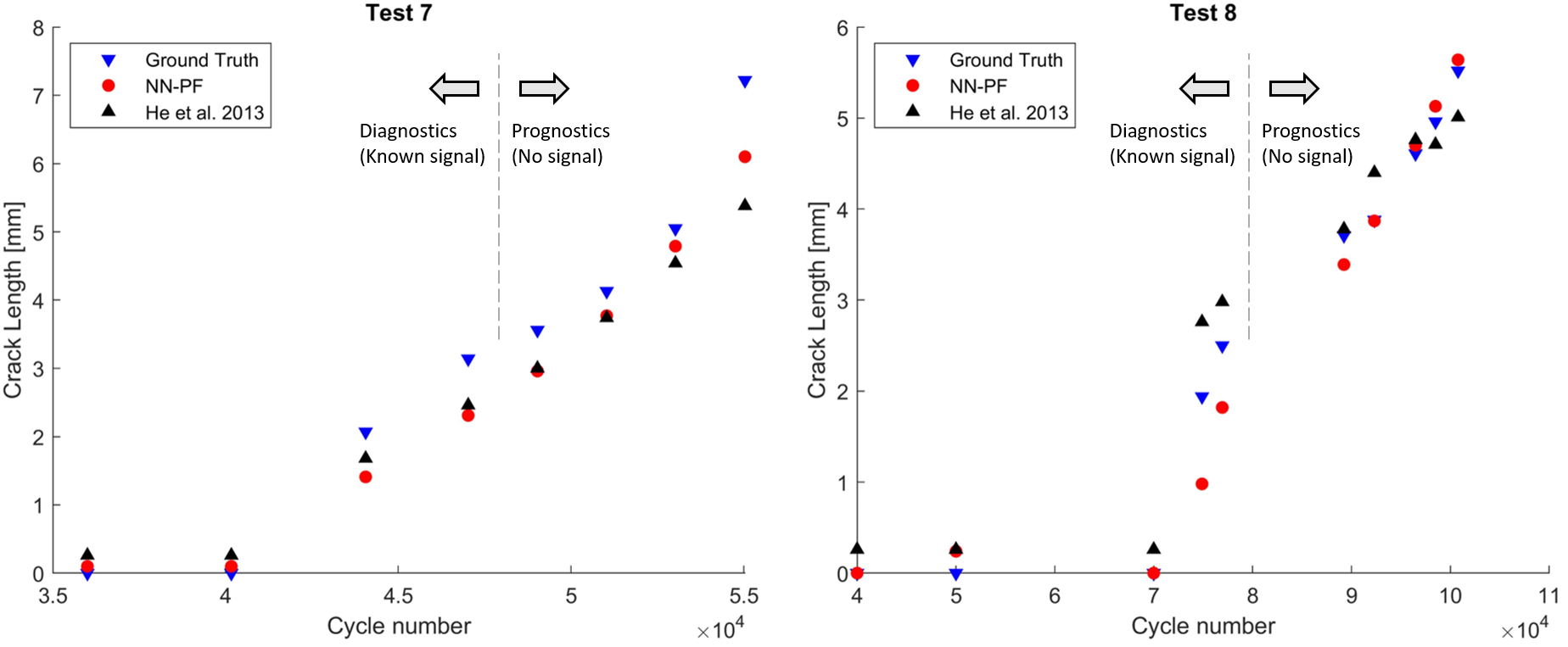}
    \caption{PF-NN framework results, ground truth, and He et al. \cite{he2013multi} predictions.}
    \label{fig:results}
\end{figure*} 

The validation tests are referred to as Test 7 and Test 8. Both tests have similar geometries as the training tests. Test 7 is performed using constant amplitude loading. In this test, eight crack lengths are to be estimated for some given cycle numbers. Ultrasonic signals are only provided for the first four cycles. Test 8 is performed under variable amplitude loading. Ten crack lengths are to be estimated for the given cycle numbers with ultrasonic signals provided only for the first five cycles. For each signal, features are extracted and fed to the neural network, which outputs an estimation of the crack length for the corresponding signal. Then, based on the NN crack length estimations, the appropriate Paris' Law parameters from the crack growth path distribution (shown in Figure \ref{fig:crack growth path}) are selected.

In Test 7, the mean value of the crack growth path shown in Figure $\ref{fig:crack growth path}$ is considered as the move function. The NN crack estimations are used by the PF to predict the crack length values at cycle numbers with available signals. These estimations are then used to update the distribution of the crack growth path shown in Figure $\ref{fig:results}$. To update the move function, the four crack lengths estimated by the PF are used to find Paris' Law parameters for Test 7. These parameters are then used to update the overall $C$ and $m$ and their confidence interval for the specimens. 
The updated moving function is then used for prognostics and crack length estimation of the remaining cycle numbers, starting from the last estimated crack length with observation (signal). Figure \ref{fig:results} shows all the PF estimations of crack length including the first four-cycle numbers (with signal) and crack length prognostics for the rest of the cycle numbers.

The NN's crack length estimations in Test 8 are closer to the upper end of the confidence interval( i.e., $5\%$ confidence). Thus, the 5\% curve's $C$ and $m$ are selected as the Paris' Law parameters for the PF's move function. NN estimations are then used as observations for PF to update the crack length values. Similar to Test 7, when no observation is available, PF is used to predict crack lengths without updates in its move function. In this case, we are not updating the move function based on a fit to Test 8 estimations due to the variable amplitude loading condition. Doing so would increase the overall uncertainty in estimated Paris Law parameters since there is no training sample with variable amplitude loading. The estimation for crack lengths for Test 8 are depicted in Figure \ref{fig:results}. 

The database in the case study has been previously used for fatigue crack estimation by He et al. \cite{he2013multi} and Wang et al. \cite{wang2018model}. In both studies, extracted features from ultrasonic signals are used to determine the parameters of proposed models. The models are based on multivariate equations by combining different features. Their approaches rely only on data-driven methods with slightly different models proposed in each study. The probability of crack detection for each proposed model is also studied in the latter research. Wang et al. \cite{wang2018model}, used physics-based modeling (Paris' Law) to estimate fatigue life (as ground truth) and evaluated the models for RUL estimation. However, to determine the distribution for the Paris' Law parameters, fatigue crack growth results of the same material reported in a different study \cite{virkler1979statistical} are used. 

\begin{figure}[h]
    \centering
    \includegraphics[width=0.45\linewidth]{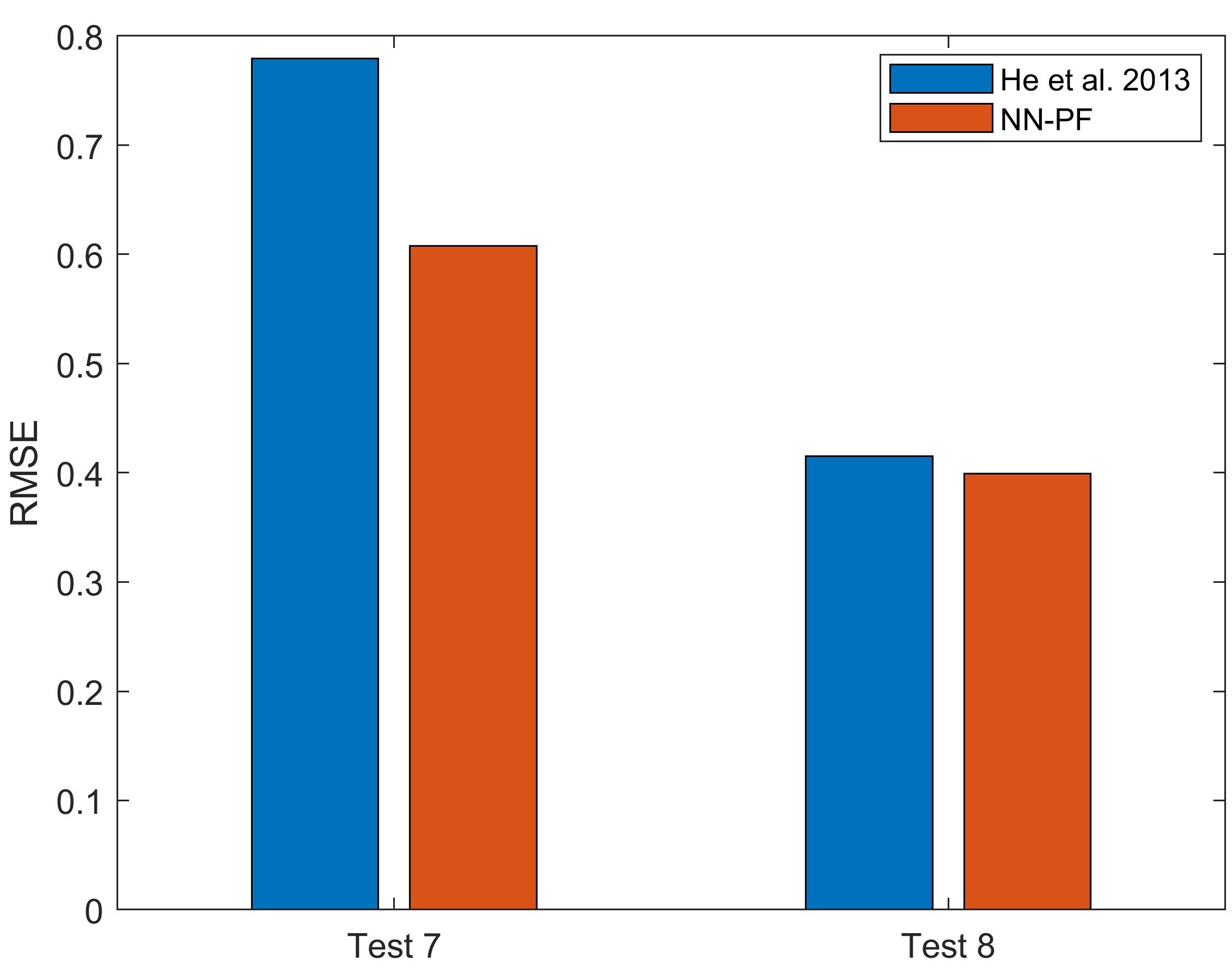}
    \caption{Root mean square error comparison of NN-PF and He et al. \cite{he2013multi}}
    \label{fig:rmse_compare}
\end{figure}

In the present study, we have used all the available data, i.e., signals, crack lengths and their associated cycle numbers to derive a hybrid physics-based and data-based method to predict crack length despite the lack of information regarding exact geometry and material. Figure $\ref{fig:results}$ shows how well our approach performs comparing to the He et al. \cite{he2013multi} and Figure $\ref{fig:rmse_compare}$ shows the root meas square error (RMSE) in each study for both tests. It can be seen that although much less information was available regarding the test conditions, material and geometry of the specimen, crack length estimation results using the proposed framework yield if not better, as good results as the other study. An accurate comparison could not be made with Wang et al. \cite{wang2018model} since they used a different validation test with slightly different training data.

\section{Conclusions}
In this paper, we have developed a hybrid \textit{Neural Network-Particle Filter} framework for crack lengths estimation based on Lamb Waves Signals and demonstrated its effectiveness in a case study. The signals in the case study were collected during multiple fatigue tests on lap-shear joint aluminum alloys. The denoising of the signals turned out to be a key factor in extracting representative features, which were then fed to a Neural Network for a first estimation of the crack length. Four features were selected and found to be sensitive to crack lengths, including the information entropy of signals which, to the best of the authors' knowledge, has not been used for the purpose of crack length estimation previously. 

Two separate time windows in signals were found to be most correlated with the actuation signal and thus used to extract the features. The features from the second time window that appears with a time delay (i.e., Time of Flight) relative to the actuation signal, were identified as the most sensitive to damage quantification. A series of potential crack growth paths with their associated probability of happening are calculated based on the available training tests. The NN crack length estimations are then used to select the appropriate crack growth path (i.e., Paris' Law parameters). 

The selected path is then considered as the move function for the particle filter. Crack length estimations from the NN were used as observation to update and adjust the PF model to decrease the uncertainty in crack length estimation further. This model allows crack propagation and RUL estimations under uncertainty, combining all the available data. Comparing to He et al. \cite{he2013multi}, this model decreased uncertainty in crack length prediction by $22\%$ in test $7$ and $4\%$ in test $8$. The proposed framework could be further improved by using the more general forms of the Paris' Law that can describe all the regions in crack growth. The comparison of the proposed framework with similar studies on the same dataset shows that fusing more of the available data, i.e., from a physics-based and data-driven approach, despite some lack in available information, can further decrease the uncertainties associated with crack length estimations and accurately predict RUL of components with minimal information on the geometry and material.

{\small
\bibliographystyle{unsrt}
\bibliography{main.bbl}

\begin{thebibliography}{10}

\bibitem{cui2002state}
Weicheng Cui.
\newblock A state-of-the-art review on fatigue life prediction methods for
  metal structures.
\newblock {\em Journal of marine science and technology}, 7(1):43--56, 2002.

\bibitem{santecchia2016review}
E~Santecchia, AMS Hamouda, F~Musharavati, E~Zalnezhad, M~Cabibbo, M~El~Mehtedi,
  and S~Spigarelli.
\newblock A review on fatigue life prediction methods for metals.
\newblock {\em Advances in Materials Science and Engineering}, 2016, 2016.

\bibitem{paris1963critical}
Pe~Paris and Fazil Erdogan.
\newblock A critical analysis of crack propagation laws.
\newblock {\em Journal of basic engineering}, 85(4):528--533, 1963.

\bibitem{he2013multi}
Jingjing He, Xuefei Guan, Tishun Peng, Yongming Liu, Abhinav Saxena, Jose
  Celaya, and Kai Goebel.
\newblock A multi-feature integration method for fatigue crack detection and
  crack length estimation in riveted lap joints using {L}amb waves.
\newblock {\em Smart Materials and Structures}, 22(10):105007, 2013.

\bibitem{yang2016probabilistic}
Jinsong Yang, Jingjing He, Xuefei Guan, Dengjiang Wang, Huipeng Chen, Weifang
  Zhang, and Yongming Liu.
\newblock A probabilistic crack size quantification method using in-situ {L}amb
  wave test and {B}ayesian updating.
\newblock {\em Mechanical Systems and Signal Processing}, 78:118--133, 2016.

\bibitem{chen2016research}
Jian Chen, Shenfang Yuan, Lei Qiu, Jian Cai, and Weibo Yang.
\newblock Research on a {L}amb wave and particle filter-based on-line crack
  propagation prognosis method.
\newblock {\em Sensors}, 16(3):320, 2016.

\bibitem{muller2017structural}
Aurelia Muller, Bradley Robertson-Welsh, Patrick Gaydecki, Matthieu Gresil, and
  Constantinos Soutis.
\newblock Structural health monitoring using {L}amb wave reflections and total
  focusing method for image reconstruction.
\newblock {\em Applied Composite Materials}, 24(2):553--573, 2017.

\bibitem{li2018damage}
Ruihua Li, Hao Li, and Bo~Hu.
\newblock Damage identification of large generator stator insulation based on
  {PZT} sensor systems and hybrid features of {L}amb waves.
\newblock {\em Sensors}, 18(9):2745, 2018.

\bibitem{yuan2017line}
Shenfang Yuan, Jian Chen, Weibo Yang, and Lei Qiu.
\newblock On-line crack prognosis in attachment lug using {L}amb
  wave-deterministic resampling particle filter-based method.
\newblock {\em Smart Materials and Structures}, 26(8):085016, 2017.

\bibitem{lu2009artificial}
Ye~Lu, Lin Ye, Zhongqing Su, Limin Zhou, and Li~Cheng.
\newblock Artificial {N}eural {N}etwork ({ANN})-based crack identification in
  aluminum plates with {L}amb wave signals.
\newblock {\em Journal of Intelligent Material Systems and Structures},
  20(1):39--49, 2009.

\bibitem{aminpour2012applying}
Hossein Aminpour, Foad Nazari, and Sara Baghalian.
\newblock Applying artificial neural network and wavelet analysis for multiple
  cracks identification in beams.
\newblock {\em International Journal of Vehicle Noise and Vibration},
  8(1):51--59, 2012.

\bibitem{sbarufatti2014numerically}
Claudio Sbarufatti, G~Manson, and K~Worden.
\newblock A numerically-enhanced machine learning approach to damage diagnosis
  using a {L}amb wave sensing network.
\newblock {\em Journal of Sound and Vibration}, 333(19):4499--4525, 2014.

\bibitem{de2015application}
A~De~Fenza, A~Sorrentino, and P~Vitiello.
\newblock Application of artificial neural networks and probability ellipse
  methods for damage detection using {L}amb waves.
\newblock {\em Composite Structures}, 133:390--403, 2015.

\bibitem{khan2015prognostics}
Faisal Khan, Omer~F Eker, Ian~K Jennions, and Antonios Tsourdos.
\newblock Prognostics of crack propagation in structures using time delay
  neural network.
\newblock In {\em 2015 IEEE Conference on Prognostics and Health Management
  (PHM)}, pages 1--6. IEEE, 2015.

\bibitem{nazarko2016damage}
Piotr Nazarko and Leonard Ziemianski.
\newblock Damage detection in aluminum and composite elements using neural
  networks for {L}amb waves signal processing.
\newblock {\em Engineering Failure Analysis}, 69:97--107, 2016.

\bibitem{cofre2019deep}
Sergio Cofre-Martel, Philip Kobrich, Enrique Lopez~Droguett, and Viviana
  Meruane.
\newblock Deep convolutional neural network-based structural damage
  localization and quantification using transmissibility data.
\newblock {\em Shock and Vibration}, 2019, 2019.

\bibitem{neerukatti2014fatigue}
Rajesh~Kumar Neerukatti, Kuang~C Liu, Narayan Kovvali, and Aditi Chattopadhyay.
\newblock Fatigue life prediction using hybrid prognosis for structural health
  monitoring.
\newblock {\em Journal of Aerospace Information Systems}, 11(4):211--232, 2014.

\bibitem{loutas2017data}
Theodoros Loutas, Nick Eleftheroglou, and Dimitrios Zarouchas.
\newblock A data-driven probabilistic framework towards the in-situ prognostics
  of fatigue life of composites based on acoustic emission data.
\newblock {\em Composite Structures}, 161:522--529, 2017.

\bibitem{shahraki2017review}
Ameneh~Forouzandeh Shahraki, Om~Parkash Yadav, and Haitao Liao.
\newblock A review on degradation modelling and its engineering applications.
\newblock {\em International Journal of Performability Engineering},
  13(3):299--314, 2017.

\bibitem{jouin2016particle}
Marine Jouin, Rafael Gouriveau, Daniel Hissel, Marie-C{\'e}cile P{\'e}ra, and
  Noureddine Zerhouni.
\newblock Particle filter-based prognostics: review, discussion and
  perspectives.
\newblock {\em Mechanical Systems and Signal Processing}, 72:2--31, 2016.

\bibitem{baraldi2013ensemble}
Piero Baraldi, Michele Compare, Sergio Sauco, and Enrico Zio.
\newblock Ensemble neural network-based particle filtering for prognostics.
\newblock {\em Mechanical Systems and Signal Processing}, 41(1-2):288--300,
  2013.

\bibitem{newman1984crack}
Jr~JC Newman.
\newblock A crack opening stress equation for fatigue crack growth.
\newblock {\em International Journal of Fracture}, 24(4):R131--R135, 1984.

\bibitem{samarasinghe2016neural}
Sandhya Samarasinghe.
\newblock {\em Neural networks for applied sciences and engineering: from
  fundamentals to complex pattern recognition}.
\newblock Auerbach publications, 2016.

\bibitem{2019PHMC82:online}
PHMsociety.
\newblock 2019 {PHM} conference data challenge – {PHM} society data
  repository.
\newblock \url{https://www.phmdata.org/2019datachallenge/}, July 2019.
\newblock (Accessed on 11/21/2019).

\bibitem{kingma2014adam}
Diederik~P Kingma and Jimmy Ba.
\newblock Adam: A method for stochastic optimization.
\newblock {\em arXiv preprint arXiv:1412.6980}, 2014.

\bibitem{KARIMIAN2020106771}
Seyed~Fouad Karimian, Mohammad Modarres, and Hugh~A. Bruck.
\newblock A new method for detecting fatigue crack initiation in aluminum alloy
  using acoustic emission waveform information entropy.
\newblock {\em Engineering Fracture Mechanics}, 223:106771, 2020.

\bibitem{sun2017lamb}
Fuqiang Sun, Ning Wang, Jingjing He, Xuefei Guan, and Jinsong Yang.
\newblock {L}amb wave damage quantification using {GA}-based {LS}-{SVM}.
\newblock {\em Materials}, 10(6):648, 2017.

\bibitem{wang2018model}
Dengjiang Wang, Jingjing He, Xuefei Guan, Jinsong Yang, and Weifang Zhang.
\newblock A model assessment method for predicting structural fatigue life
  using {L}amb waves.
\newblock {\em Ultrasonics}, 84:319--328, 2018.

\bibitem{sinclair2010enhancement}
AN~Sinclair, J~Fortin, B~Shakibi, F~Honarvar, M~Jastrzebski, and MDC Moles.
\newblock Enhancement of ultrasonic images for sizing of defects by
  time-of-flight diffraction.
\newblock {\em NDT \& e International}, 43(3):258--264, 2010.

\bibitem{nath2010reliability}
SK~Nath, Krishnan Balasubramaniam, CV~Krishnamurthy, and BH~Narayana.
\newblock Reliability assessment of manual ultrasonic time of flight
  diffraction ({TOFD}) inspection for complex geometry components.
\newblock {\em NDT \& E International}, 43(2):152--162, 2010.

\bibitem{sedlacek2005digital}
Milos Sedlacek and Michal Krumpholc.
\newblock Digital measurement of phase difference-a comparative study of dsp
  algorithms.
\newblock {\em Metrology and Measurement Systems}, 12(4):427--448, 2005.

\bibitem{shannon1948mathematical}
Claude~Elwood Shannon.
\newblock A mathematical theory of communication.
\newblock {\em Bell system technical journal}, 27(3):379--423, 1948.

\bibitem{virkler1979statistical}
Dennis~Andrew Virkler, Brnm Hillberry, and PK~Goel.
\newblock The statistical nature of fatigue crack propagation.
\newblock {\em Journal of Engineering materials and Technology},
  101(2):148--153, 1979.

\end{thebibliography}
}

\end{document}